\documentclass[11pt]{article}
\usepackage{graphicx,color}
\usepackage{rotating}
\usepackage{epsfig,graphics,rotate,color}
\usepackage{wrapfig}
\advance \topmargin by -\headheight
\advance \topmargin by -\headsep
\evensidemargin \oddsidemargin
\usepackage[T1]{fontenc}
\usepackage{times}
 \usepackage{hyperref}

\begin{document} 
\begin{center}
\Large{Next-Generation Liquid-Scintillator-Based Detectors: \\ Quantums Dots and Picosecond Timing}\\[0.25cm]
\large{Lindley Winslow,  \today}\\[0.25cm]
\end{center}

Liquid-scintillator-based detectors are a robust technology that scales well to large volumes. The kiloton-scale detectors KamLAND\cite{Gando:2010aa} and Borexino\cite{borexinobe7} have been very successful, each providing key evidence that neutrinos oscillate and therefore have mass. Now that neutrino mass is an established fact, the quest is on to determine the nature of that mass and answer the deeper question of the Majorana or Dirac nature of the neutrino. The observation of the rare nuclear process neutrinoless double-beta decay $(0\nu\beta\beta)$ would prove that the neutrino is Majorana. In the search for this rare process, the kiloton-scale scintillator detectors are already contributing with KamLAND-Zen\cite{KZ0nu,KZMaj,KZ2nu}  or will be soon with SNO+\cite{snop}. 

The technology is known to suffer from poor energy resolution compared to germanium detectors, tellurium bolometers, and liquid or gas time projection chambers; however, this can be compensated for by increasing the mass of the double-beta decay isotope.  The required increase in isotope mass is relatively straightforward compared to the other technologies. Scintillator-based detectors also have an advantage due to the significant self-shielding of backgrounds and the ability to do position reconstruction, which allows further reduction of the backgrounds. 

The minimum goal for the next generation of liquid-scintillator-based detectors is an increase in the energy resolution due to a combination of increased photocathode coverage and increased light yield with reduced attenuation length of the scintillator liquid.  A more ambitious goal is to get a second signal out of the detector, ideally the directionality of the betas in addition to their total energy.  Reconstructing the direction of the betas would allow us to differentiate light Majorana neutrino exchange from more exotic mechanisms of $0\nu\beta\beta$\cite{angcorr}, and even if the direction of the betas cannot be fully reconstructed, any event topology signal will be valuable for rejecting backgrounds.

The ability to reconstruct direction in these detectors will come from the ability to distinguish the directional Cerenkov light from the abundant isotropic scintillation light. The Cerenkov light which is not absorbed and reemitted by the scintillator and retains its directional information is at longer wavelengths and therefore arrives approximately 1~ns before the scintillation light. As an example, for a 1~MeV $\beta$ we expect $\sim$60 Cerenkov photons, which means that close to 100\% photocathode coverage with $\sim$100~ps timing is needed. The photocathode coverage criteria is similar to that for Daya Bay-II\cite{dayabayII}, and the timing criteria matches that for the device being developed by the LAPPD collaboration\cite{LAPPDSum,LAPPDTDR}.

The other variable in the design of a scintillator-based detector is the chemical composition of the liquid. It could be that the addition of quantum dots, nanocrystals with the unique optical property that they absorb all light shorter than a characteristic wavelength and reemit within a narrow peak around this wavelength, to the scintillator may enhance the discrimination of the longer wavelength directional Cerenkov light relative to the isotropic scintillation light.  Quantum dots can be made from such elements as Cd, Te, Zn, and Se and would therefore suspend the needed $0\nu\beta\beta$ target isotope into the liquid. A Cd-doped scintillator would also have application for antineutrino experiments using inverse beta decay especially for non-proliferation applications. Preliminary work characterizing quantum-dot-doped liquid scintillator is reported in Ref.~\cite{qdot}.

The combination photodetector and scintillator development represents a rich R\&D program for next few years. The first opportunity to implement these ideas would be in KamLAND. A proposal for upgrades to the detector would go to the Japanese funding agencies for the first time in 2015. The schedule for KamLAND running is fairly full, and the schedule of the refurbishment will be set by the physics goals and needs of each run; however, refurbishment should be completed by 2020. The current cost of the KamLAND liquids and PMTs with electronics are \$8M and \$6.6M respectively, so the replacement, plus enhancements, put the total cost of such a project at $\sim$\$20M. 
\bibliography{NSF_Winslow2012} 
\end{document}